# Chaotic Encryption Applied to Optical Ethernet in Industrial Control Systems

Adrián Pérez-Resa ![ORCID], Miguel Garcia-Bosque, Carlos Sánchez-Azqueta, and Santiago Celma

*Abstract*—In the past decades, Ethernet has become an alternative technology for the field buses traditionally used in industrial control systems and distributed measurement systems. Among different transmission media in Ethernet standards, optical fiber provides the best bandwidth, excellent immunity to electromagnetic interference, and less signal loses than other wired media. Due to the absence of a standard that provides security at the physical layer of optical Ethernet links, the main motivation of this paper is to propose and implement the necessary modifications to introduce encryption in Ethernet 1000Base-X standard. This has consisted of symmetric streaming encryption of the 8b10b symbols flow at physical coding sublayer level, thanks to a keystream generator based on chaotic algorithm. The overall system has been implemented and tested in an field programmable gate array and Ethernet traffic has been encrypted and transmitted over an optical link. The experimental results show that it is possible to cipher traffic at this level and hide the complete Ethernet traffic pattern from passive eavesdroppers. In addition, no space overhead is introduced in data frames during encryption, achieving the maximum throughput.

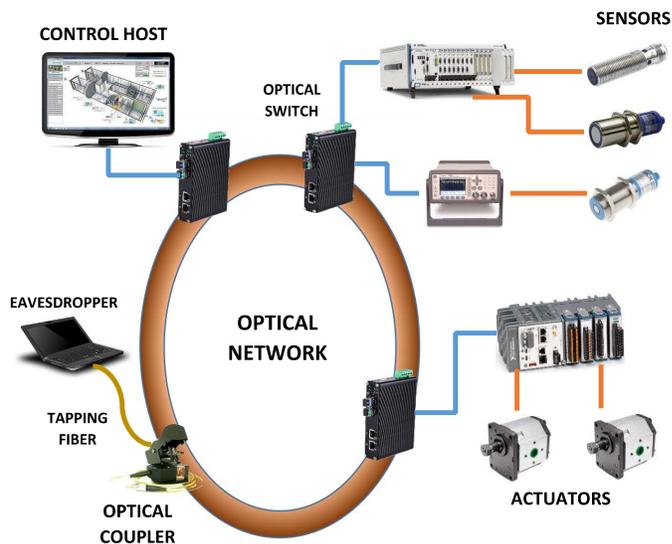

Fig. 1. Simple example of a distributed control and data acquisition network and an eavesdropper attack with a tapping fiber.

## I. INTRODUCTION

IN RECENT decades, Ethernet has expanded widely in industrial control systems and critical infrastructures, replacing the traditional communication field buses [1], [2]. Ethernet has proved to be an efficient technology to create distributed acquisition systems such as supervisory control and data acquisition networks. At present, this kind of networks not only supports measurement and control systems but also data traffic for surveillance video security and future Internet of Things (IoT) applications. Both, distributed measurement platforms with integrated Ethernet ports and industrial network equipment up to 1-Gb/s rates and beyond, are available from many vendors. A simple scheme of a distributed control and data acquisition network is shown in Fig. 1.

Manuscript received October 2, 2018; revised January 18, 2019; accepted January 21, 2019. The work of M. Garcia-Bosque was supported in part by MINECO-FEDER under Grant TEC2014-52840-R and Grant TEC2017-85867-Rand in part by FPU Fellowship under Grant FPU14/03523. The Associate Editor coordinating the review process was Huang-Chen Lee. *(Corresponding author: Adrián Pérez-Resa.)*

The authors are with the Electronic and Communications Engineering Department, Zaragoza University, 50009 Zaragoza, Spain (e-mail: aprz@unizar.es; mgbosque@unizar.es; csanaz@unizar.es; scelma@unizar.es).

Color versions of one or more of the figures in this paper are available online at http://ieeexplore.ieee.org.

Digital Object Identifier 10.1109/TIM.2019.2896550

Optical Ethernet is widely used in industrial environments. Optical fiber has some advantages over other wired methods, such as higher bandwidth, less signal losses, and more immunity to electromagnetic interference. In addition, as it does not emit any radiation, it is safer than wireless systems that are more exposed to eavesdropping.

However, vulnerability and threat analysis in the physical layer of optical systems is critical to guarantee secure communications [3], [4]. One of the most important attacks is the splitting attack. It is normally used for either eavesdropping or signal degradation and can be easily performed with tapping techniques. At present, low-cost methods for intercepting the optical signal through fiber coupling devices and electrooptical converters are available without the need to perceptibly interfere in communications [5]. To avoid or detect eavesdropping, encryption and intrusion detection systems have been proposed as solutions [3].

In a layered communication model, encryption methods can be implemented at different communication levels. It depends on the communication layer where confidentiality is needed. In the particular case of industrial Ethernet, solutions are usually proposed for network and transport layers (layers 3 and 4), such as IPsec or transport layer security protocols [6], [7]. Other solutions are proposed for the data link layer (layer 2), such as MACsec standard [8].

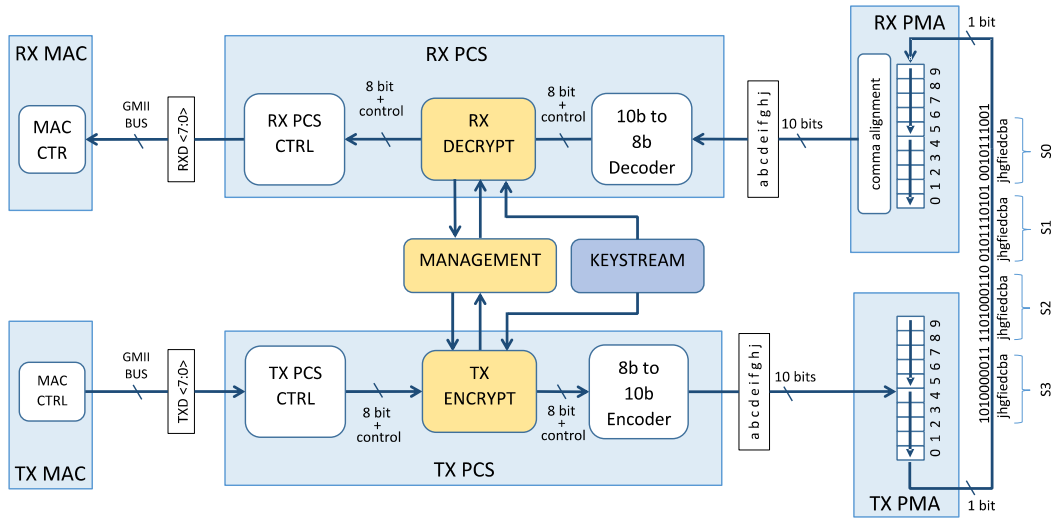

Fig. 2. PCS structure with the proposed encryption function PHYsec included.

Regarding physical layer encryption in optical communications, there are some solutions for protocols such as optical transport network [9]. However, they are only oriented to telecommunication networks. One advantage of encryption at the physical layer is that it can be performed at a line-rate introducing a very low latency. In the particular case of industrial Ethernet networks, high level of determinism is usually needed, and for certain applications, there can also be strict requirements for network delay and jitter [10]. A physical layer encryption mechanism could be useful for encrypting traffic of real-time protocols such as EtherCAT or PROFINET IRT [10], [11], enabling encryption to be performed at line rate.

As 1000Base-X standard is one of the most widely used physical layer standards at 1 Gb/s, the main motivation of this paper is to propose and implement an encryption method suitable for the physical coding sublayer (PCS) of this standard. As the main coding in 1000Base-X is the well-known 8b10b encoding, a stream cipher that preserves the format of this codification is needed.

In general, two approaches are usually used to build stream ciphers: *ad hoc* stream ciphers, as those proposed in [12], and secure blockciphers, such as advanced encryption standard (AES), working in a specific operation mode, e.g., counter mode (CTR). These two approaches are focused on plaintext in binary format, which are not suitable for our purposes. Among different *ad hoc* stream ciphers solutions, there are those based on chaotic algorithms. The characteristics of these are very similar to those that a good cryptographic algorithm should have. They have been analyzed from several points of view [13], [14] and have given rise to new encryption proposals related with several areas, such as IoT applications [15], communications [16], or image encryption [17], [18]. On the other hand, chaotic encryption has been applied with different technologies, using analog and digital electronic circuits [19], [20] or optical and electrooptical mechanisms [21], [22].

To encrypt data preserving its format with a stream cipher, a solution was initially proposed in [23], where the keystream generator was based on a format preserving encryption (FPE) blockcipher working in CTR mode.

This paper is an extension of [24], where a new alternative to [23] is proposed. In this paper, this encryption method has been named PHYsec and the keystream generator is now based on a chaotic stream cipher.

An electronic implementation for the chaotic algorithm has been carried out in a digital programmable platform as in [23]. Thanks to the proposed structure, it is possible to save hardware resources while reaching the same encryption throughput. Moreover, the power consumption is reduced by more than 50%.

In addition, this paper deals with another important issue relative to the encryption system such as the keystream synchronization, a topic that is not addressed in [23].

This paper is organized as follows. Section II explains the overall structure of the encryption mechanism in the PCS sublayer. Sections III and IV deal with the encryption operation for PHYsec and the synchronization mechanism between TX (transmitter) and RX (receiver), respectively. Section V explains the keystream generation and some considerations about the key space. Section VI details the system implementation and test results. Finally, Section VII summarizes the conclusions obtained in this paper.

## II. PCS LAYER ENCRYPTION

The physical layer or PHY is responsible for carrying out the lower level functions in the transmission. It defines the electrical, mechanical, and functional specifications to activate, maintain, and deactivate the physical link. In the case of Ethernet standards, it is usually divided into three other sublayers with different functionalities: PCS, physical medium attachment (PMA), and physical medium dependent. PCS layer performs functions such as autonegotiation, link establishment, 8b10b data encoding, symbol synchronization, clock rate adaptation, and so on.

The basic structure of the PCS layer and its interface with the medium access control (MAC) and the PMA levels are shown in Fig. 2. The coding function is separated into encoder and decoder blocks while the rest of the functions of the PCS layer would reside in the RX_PCS_CTRL and TX_PCS_CTRL modules.

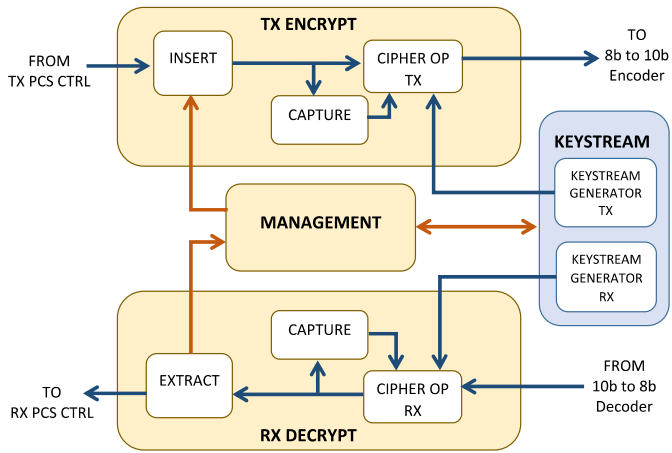

Fig. 3. Encryption infrastructure for PHYsec function.

Unlike other encryption mechanisms such as IPsec or MACsec, where the data packets are encrypted individually, in this paper, the 8b10b symbol flow is directly and continuously encrypted. One advantage of this method is that no extra data field is added to packets, which would reduce the overall data throughput. In addition, by encrypting the complete 8b10b symbol flow, it also achieves to mask the activity or data traffic pattern. This is because by encrypting at 8b10b symbol level, control symbols and ordered sets are also encrypted, such as packet start and end symbols or idle sets in the interframe gap (IFG). Some important properties of the 8b10b encoder are the transition density and the short run length introduced in the encoded serial data. These are necessary to facilitate the operation of the remote clock and data recovery (CDR) circuit. To preserve these properties, it is proposed to perform the encryption operations at the input of the encoder and decryption at the output of the decoder, as shown in Fig. 2.

One advantage of performing the encryption preserving the coding properties and maintaining the rest of physical layer features untouched is that the physical layer encryption maintains compatibility with the subsequent hardware elements or medium-dependent circuitry. For example, commercial optical modules as small form-factor pluggable (SFP) and electronic circuits as CDR or serializer/deserializer (SERDES) for 1000Base-X standard would be also compatible with the proposed encryption method.

In the proposed system, encryption is carried out by a symmetrical stream cipher, as it fits well with a continuous data flow such as 8b10b symbol transmission. We divide the system into three functional parts. The first consists of the encryption and decryption blocks, which includes the mathematical stream cipher operation. The second is the synchronization mechanism between TX and RX. Finally, the third is the keystream generator used in the cipher function. These parts are explained in Sections III, IV, and V, respectively.

## III. ENCRYPTION AND DECRYPTION OPERATIONS

The stream cipher operation is performed in the CIPHER_OP_TX and CIPHER_OP_RX modules in Fig. 3. The data set to be encrypted is a limited set, since the valid 8b10b symbols are composed of 256 data symbols plus 12 control symbols, a total of 268 symbols which do not generate code errors.

Nevertheless, one of the 12 control symbols (/K28.7/) is not used for standard data communication [25], and to avoid its accidental generation in the encryption of any 8b10b symbol, it has been excluded from the encryption symbol mapping, explained in the following. Therefore, the number of valid symbols is 267.

As the encryption is performed before the encoding and it is necessary to preserve the coding properties, the encrypted data must be valid 8b10b symbols inside the 267 possible symbols. To achieve this goal, the symbols are mapped to an integer value in the range of 0–266. After the mapping, stream cipher operation is performed. This operation consists of a modulo-267 addition between the mapped symbols and the keystream, which also takes values uniformly distributed between 0 and 266. Once the cipher operation is done, the resulting values are reverse-mapped to the corresponding new 8b10b symbol.

The decryption process is the same, only that the decryption operation is a modulo-267 subtraction. In Fig. 4, the structure of the module CIPHER_OP_TX and CIPHER_OP_RX is shown.

## IV. ENCRYPTION SYNCHRONIZATION

### A. Management Module

MANAGEMENT module in Fig. 3 is the part of the system that configures, controls, and reports the encryption status between both PHYs. It implements the initial synchronization procedure and collects the alarms relative to the synchronization status, by which the user can perform the necessary actions for achieving a coherent communication between receiver and transmitter.

For example, in case of a mismatch between encryption status (one PHY encrypting and the other not), or a bad synchronization status (misaligned keystream generators between remote PHYs), several alarms can be triggered and latched in the MANAGEMENT module, which reports them to the user, thanks to the field-programmable gate array (FPGA) debug system. The system is able to detect the loss of synchronization $2.136\ \mu s$ after the misalignment is produced. After this notification, the user is able to stop keystream generators and restart encryption synchronization procedure in a way that a new coherent encryption status between the two PHYs is achieved. In the long term, under normal working conditions, we could consider that the probability of a synchronization loss is negligible, as the implemented system uses commercial components compatible with the standard. For example, SFP modules, optical fiber, or the transceiver circuitry. The standard guarantees a very low bit error rate, under $10^{-12}$, and the ability of clock recovering with a clock tolerance of $\pm 100$ ppm.

In order to maintain the concordance with the standard, communication between the MANAGEMENT modules of remote PHYs has been planned to be based on control messages formatted like 1000Base-X ordered sets. The insertion/extraction of the new and future control messages in the 8b10b data flow is performed by MANAGEMENT module,

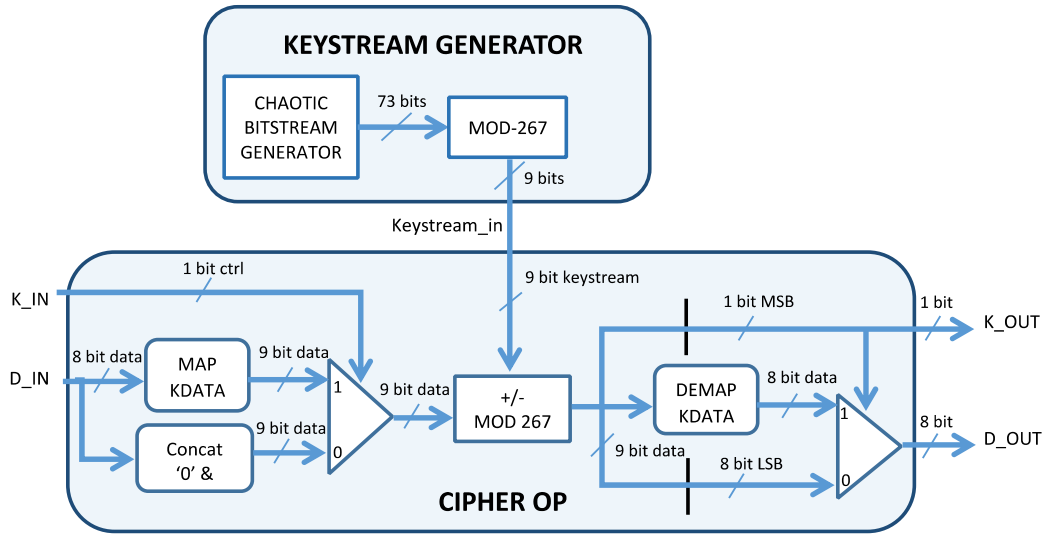

Fig. 4.   Stream cipher operation performed in CIPHER_OP module next to one of the keystream generators (RX or TX).

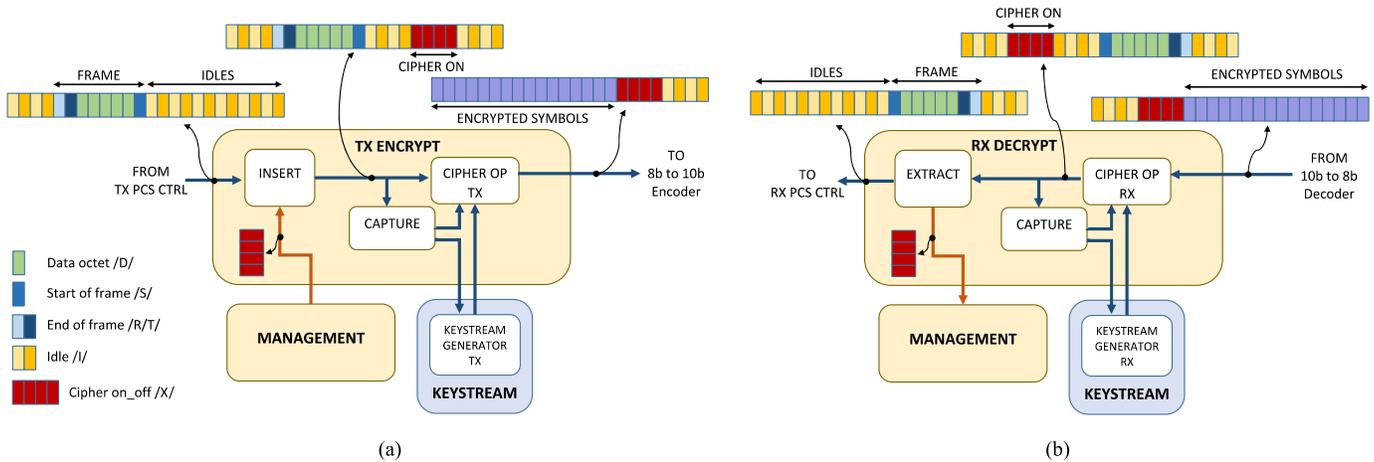

Fig. 5.   Initial synchronization procedure, based on the proposed /X/ ordered set. (a) Transmission. (b) Reception.

thanks to INSERT and EXTRACT blocks in Fig. 3, and it is explained in Section IV-B.

### B. Initial Synchronization Procedure

To achieve encryption synchronization with a stream cipher, it is necessary that the keystream sequence generated at RX side is exactly aligned with the incoming keystream sequence embedded into the ciphertext and generated at TX side. The initial synchronization procedure is in charge of this mechanism. For that purpose, MANAGEMENT module in Fig. 3 performs the insertion and extraction of a control sequence into the 8b10b symbol stream to indicate the remote PHY when to start/stop its decryption process. This sequence is used for both the activation and deactivation of the encryption and decryption.

The format of the implemented control sequence is similar to an ordered set one. In Table I, this sequence is called /X/ and it is shown along with those that already exist in the standard. The first character in the new set is /K28.1/ and as /K28.5/, it includes the comma sequence.

The basic operation of this procedure is described next and is shown in Fig. 5(a) and (b). In Fig. 5(a), the CIPHER_OP_TX module, responsible for performing the encryption operation, is initially disabled, allowing the 8b10b symbols to be transparently passed from the TX_PCS_CTRL to the 8b10b encoder. In order to start encrypting the 8b10b symbol stream, the MANAGEMENT module acts on the INSERT block to insert a /X/ encryption start message in the 8b10b symbol flow. As shown in Fig. 5(a), this message is represented as a set of four symbols and it is inserted in the line by replacing idle ordered sets with the octets that form the message itself. Before passing through the CIPHER_OP_TX module, /X/ is detected by the CAPTURE module. Then the CAPTURE module enables CIPHER_OP_TX and the keystream generator (KEYSTREAM_GENERATOR TX), which initiates the encryption process after /X/ has been transmitted. As shown in Fig. 5(a), at the output of the CIPHER_OP_TX module, every 8b10b symbol after /X/ is encrypted, including idle sets and complete Ethernet frames.

TABLE I
IEEE 802.3 ORDERED SETS [25] AND PROPOSED ORDERED SET /X/

| Code | Ordered_Set | Number of Code-Groups | Encoding |
|---|---|---|---|
| /C/ | CONFIGURATION | | Alternating /C1/ and /C2/ |
| /C1/ | Configuration 1 | 4 | /K28.5/D21.5/Config_Reg |
| /C2/ | Configuration 2 | 4 | /K28.5/D2.2/Config_Reg |
| /I/ | IDLE | | Correcting /I1/, Preserving /I2/ |
| /I1/ | Idle 1 | 2 | /K28.5/D5.6/ |
| /I2/ | Idle 2 | 2 | /K28.5/D16.2/ |
| | ENCAPSULATION | | |
| /R/ | Carrier_extend | 1 | /K23.7/ |
| /S/ | Start_of_Packet | 1 | /K27.7/ |
| /T/ | End_of_Packet | 1 | /K29.7/ |
| /V/ | Error_Propagation | 1 | /K30.7/ |
| **ENCRYPTION** | | | **Enable/Disable** |
| **/X/** | **Cipher_on_off** | **4** | **/K28.1/D21.5/D21.2/D21.2/** |

At the receiver, as shown in Fig. 5(b), the module CIPHER_OP_RX is in charge of performing the decryption operation. As in the transmitter, it is initially inactive, enabling 8b10b symbols to be transparently passed from the 8b10b decoder to the controller RX_PCS_CTRL. When the CAPTURE module detects /X/ in reception, decipher module (CIPHER_OP_RX) and the keystream generator (KEYSTREAM_GENERATOR RX) are enabled, starting to decrypt the 8b10b stream after the /X/ set. Subsequently, the control message /X/ is extracted from the data stream in the EXTRACT module, which replaces it with idle ordered sets in the 8b10b symbol flow (reverse operation of the INSERT module).

Thanks to this procedure, the beginning of the ciphertext is detected at the receiver allowing to generate a keystream sequence that is aligned with the incoming data, then performing decryption properly.

The way to insert set /X/ is shown in Fig. 6. INSERT module contains a buffer where the MANAGEMENT module can write messages while there is enough space. These messages can only be read from the buffer when the INSERT_MACHINE block in Fig. 6 indicates that it is possible to do so.

The INSERT_MACHINE block monitors a pipeline where the 8b10b symbol stream passes through. Initially, the output of the INSERT module is selected as the output of the pipeline. When a number of idle ordered sets, equivalent to the size of the next message to be read, is detected inside the pipeline, the output of INSERT module is switched to the data coming from the message buffer. Then the pending message is read and the pipeline is emptied from the stored idles. Once the message has been transmitted, the output of INSERT module switches to the pipeline output again.

## V. KEYSTREAM GENERATOR

The keystream generator is responsible for calculating pseudorandom numbers to carry out the encryption of the signal. As the encryption operation consists of a modulo-267 addition, the values of the pseudorandom sequence must be between 0 and 266. In order to build that sequence two steps are carried out in each keystream generator (TX and RX), first, a pseudorandom bitstream is generated based on a chaotic algorithm, which is considered the core of this stream cipher. Second, modulo-267 operation is applied to the output of the chaotic algorithm. The result is a pseudorandom sequence composed of 9-bit words with values between 0 and 266. The keystream generator structure is shown in Fig. 4, where the CHAOTIC_BITSTREAM_GENERATOR generates the pseudorandom bitstream and MOD-267 block carries out the modulo operation.

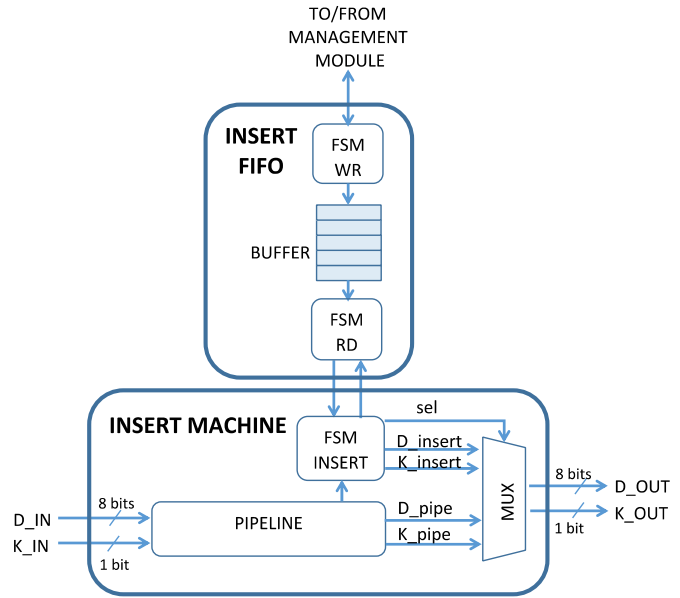

Fig. 6. INSERT module. Several finite-state machines (FSMs) govern this module. FSM_WR and FSM_RD control the write and read of message buffer and FMS_INSERT the message insertion on the line.

### A. Modulo Operation

The modulo operation must be applied to words with 9 bit width or more to result in a 9-bit-width output. When doing such operation, a bias is introduced in the distribution of the resulting number sequence. According to the National Institute of Standards and Technology (NIST) recommendation [26], in order to make this bias negligible, the input width of the modulo operation shall be at least 64 bits longer than the output. As in our case 9-bit numbers are necessary, the pseudorandom bitstream generator output and modulo operator input have to be at least 73 bits width.

The implementation of modulo-267 operation has been based on [27], which presents a high-speed hardware implementation for a generic operation "$x \bmod z$" that can be fragmented into a pipeline of $n - m + 1$ stages, where $x$ is represented by $n$ bits and $z$ by $m$ bits. In our case, $z$ has been taken as a constant value equal to 267. According to this, the resulting hardware structure should have 65 stages as shown in Fig. 7. However, by implementing an overclocked structure, this number can be reduced to 33.

### B. Pseudorandom Bitstream Generator

The study of chaos and the systems derived from it have aroused interest as possible design solutions of new cryptographic systems. From a theoretical point of view, a good cryptographic algorithm should have properties such as confusion, diffusion, and sensitivity to changes in plaintext and secret key. Apart from that, the most well-known characteristics of chaotic systems are the high sensitivity to the initial conditions or control parameters (also called "butterfly effect") and the unpredictable pseudorandom orbits generated by their algorithms.

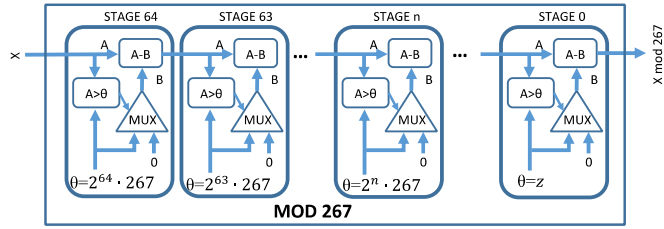

Fig. 7. Modulo 267 hardware.

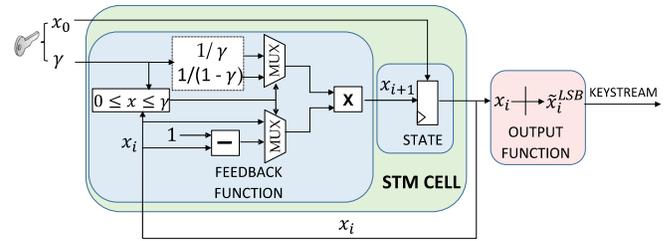

Fig. 8. STM block diagram.

Among chaos-based cryptosystems, there are two main approaches for their design. On the one hand, analog chaotic cryptosystems based on chaos synchronization, where two chaotic systems can synchronize under the driving of scalar signals sent from one system to another. These kinds of cryptosystems are usually implemented in the analog domain and need very accurate components to ensure information recovery.

On the other hand, a second approach is the discrete chaos-based cryptosystems, where one or more chaotic maps are implemented using finite precision and they do not depend on chaos synchronization at all. Many hardware implementations have been proposed for a wide variety of chaotic maps such as in [28], [29], or [30] where the well-known modified logistic map, Hennon map, and Bernoulli map, respectively, were built. Although there are analog solutions that have been digitized, as in [20], the discrete chaos-based approach is more suitable for digital hardware platforms as FPGAs.

In this paper, for generating the pseudorandom bitstream, a variant of the chaotic map called skew tent map (STM) has been used. This chaotic map has already been extensively studied in [31] and [32], based on the previous work by our group.

The equations of this map are shown in the following equation:

$$f(x_i) = x_{i+1} = \begin{cases} x_i/\gamma, & x_i \in [0, \gamma] \\ (1-x_i)/(1-\gamma), & x_i \in (\gamma, 1] \end{cases} \quad (1)$$

where its control parameter and initial state are $\gamma$ and $x_0$, respectively, and their values are included in the interval (0, 1). One important characteristic of STM is that it is a piecewise linear map and has positive Lyapunov exponent for any selected value of $\gamma$ and $x_0$. This fact can satisfy always the chaoticity of this map [13] and makes it more adequate than others for cryptographic applications. In Fig. 8, the hardware block diagram of the basic STM algorithm is shown.

In this paper, a fixed-point implementation of the STM algorithm has been carried out. A typical problem of this kind of implementations is that due to the finite word length, degradation is produced in the dynamical properties of the chaotic map. When the STM is implemented with a presicion of $n$ bits, there are only $2^n$ possible values for each $x_i$, giving a maximum period for the keystream of $2^n$ possible samples. In spite of this, cycle lengths are found empirically to be much shorter. Owing to this fact, usually some randomness test applied to the chaotic sequences fail. One recommended solution for increasing this period is the use of a pseudorandom number generator to perturb the chaotic orbits of this kind of algorithms [33].

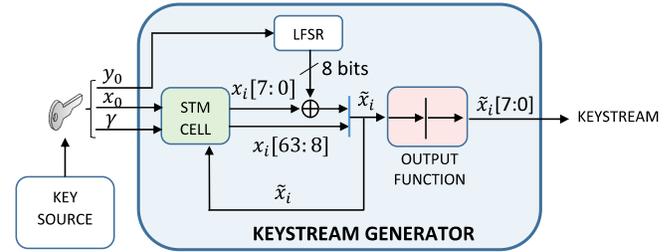

Fig. 9. Basic keystream generator.

For this purpose, the structure in Fig. 8 has been improved by using a linear feedback shift register (LFSR), obtaining a significant increase in the keystream period and, therefore, better properties as a pseudorandom sequence [32], [34]. According to [32], by selecting an LFSR with 61 steps, it is possible to guarantee an enough length for the chaotic sequence that makes it pass the randomness tests.

Fig. 9 shows the structure of the basic keystream generator used in this paper. A 61-step LFSR has been added to the STM module. The cryptographic key used to configure this basic generator will consist of the initial state and control parameter $(\gamma, x_0)$ of STM cell and the initial state of the LFSR $(y_0)$.

The STM cell has been implemented with an internal 64-bit state whose output is the vector $x_i[63 : 0]$. The improvement related to LFSR is to perform the XOR operation between the least significant 8 bits of LFSR output and the STM cell before applying the generator output function. The state returned to the STM cell will no longer be the previous state calculated by it, but a new state $\tilde{x}_i$ to which a small noise generated by the LFSR has been added.

The generator output function in Fig. 9 is to take the 8 least significant bits of $\tilde{x}_i$. Since the output bus of the keystream generator must be 73 bits width, a bank of basic keystream generators has been built. The bank consists of eight 8-bit-width generators and one 9-bit-width generator; their outputs are concatenated to give a 73-bit output. The final structure for the keystream generator is shown in Fig. 10.

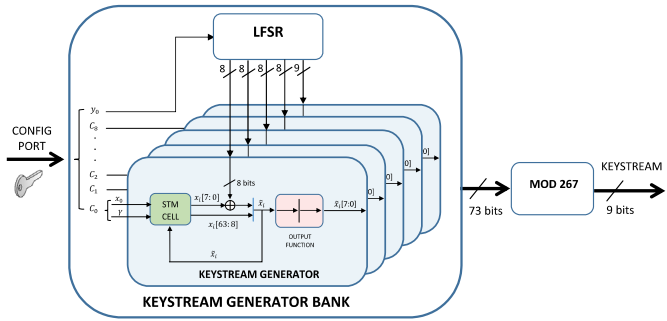

Fig. 10.    Keystream generator.

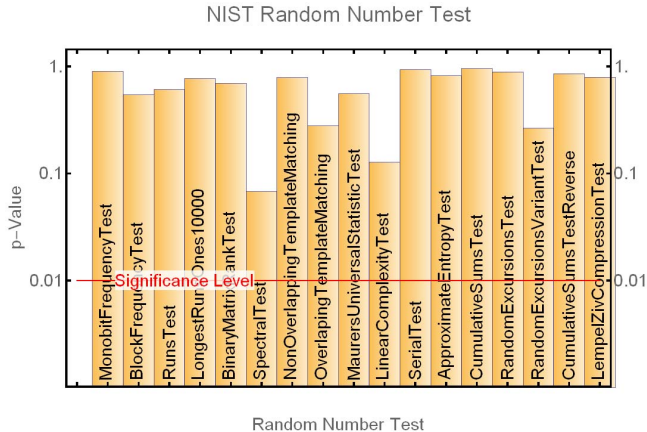

Fig. 11.    NIST test results for a bitstream generated using STM-LFSR algorithm.

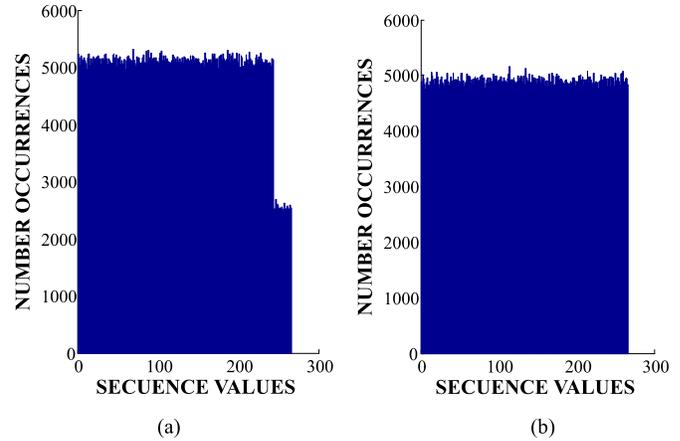

Fig. 12.    Histogram of output keystream (a) Without applying NIST recommendation and (b) Applying it.

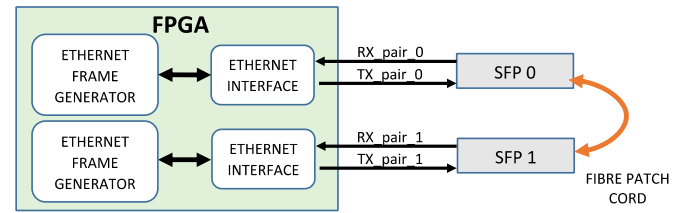

Fig. 13.    Test setup scheme.

### C. Keystream Output Analysis

One of the main requirements of a secure stream cipher is that its keystream should be undistinguishable from a truly random sequence. On the one hand, it is necessary to test the randomness of the chaotic bitstream, and on the other hand, it is necessary to check that the bias introduced by module 267 operator is negligible.

In order to test the randomness of chaotic bitstream, sequences generated by the basic generator in Fig. 9 were subjected to the NIST SP 800-22 battery of tests [35]. The generator passed these tests, and an example of the results for a particular sequence is shown in Fig. 11.

In order to check the uniformity between 0 and 266 of the resulting keystream after modulo operation, histograms for different sequences were obtained. In Fig. 12, two cases are shown for a sequence of 1 310 720 values, one of them [Fig. 12(b)] is the histogram obtained when NIST recommendation [26] is applied and the input to module 267 operator is 73 bit width. The other case [Fig. 12(a)] is when only a 9-bit-width input is used in the module operation. Clearly, a bias appears in the second case, but it seems negligible in the first.

Moreover, chi-square goodness of fit test has been used to determine if the resulting keystream obtained when applying [26] comes from a uniform distribution between 0 and 266. The test has not rejected this hypothesis at 5% of significance level. However, this kind of tests for uniform distributions does not guarantee the randomness for the final keystream.

The NIST battery of test is suitable for binary streams of data; however, it is not directly applicable to nonbinary data sets when they are not power of two. As far as we are concerned, there is not a standardized set of test for checking randomness of nonbinary sources.

In this paper, some tests suitable for nonbinary sources described in [36] have been applied. Particularly, frequency test, serial test, and poker test were used successfully. As for frequency and serial test, the chi-square goodness of fit test was successfully passed to sequences of 3 and 15 millions of tuples, respectively. Regarding the poker test, also the chi-square goodness of fit test was carried out by using the five categories described by Knuth [36].

### D. Key Space

In the proposed system, the key space size of a single cell would be given by all the possible values of $\gamma$ and $x_0$, with 64 bits each one, and the initial 61-bit-length LFSR state, $y_0$. This gives a total key space size of $2^{64+64+61} = 2^{189}$. Some guidelines recommend that in order to be secure for the coming years, the key space size should be greater than $2^{112}$ or $2^{128}$ [37], [38]. Therefore, the key space size of this algorithm is big enough to prevent both key reusing as well as brute-force attacks.

## VI. SYSTEM IMPLEMENTATION

### A. System Description

The complete system has been implemented in a Xilinx Virtex 7 FPGA. In the setup for test, the FPGA has been connected to two SFP modules capable of transmitting at a rate of 1.25 Gb/s at 850 nm over multimode fiber.

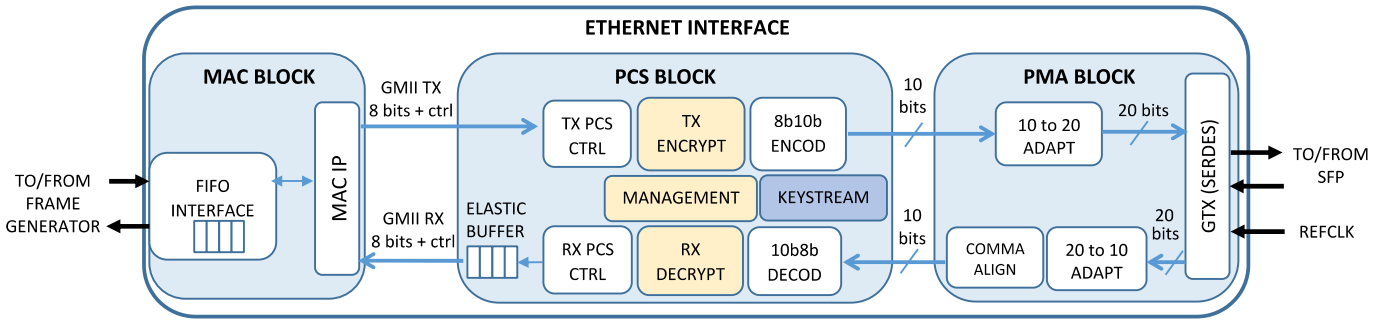

Fig. 14. Scheme of FPGA implementation for Ethernet interface with the encryption function. TX and RX PCS datapaths work with different clock domains. In TX datapath, the system clock is used. However, for the RX datapath, is necessary to use the recovered clock from PMA block. This clock is used on the RX side of the PCS block until the elastic buffer. Elastic buffer is used for rate adaptation between MAC and RX interface of PCS block.

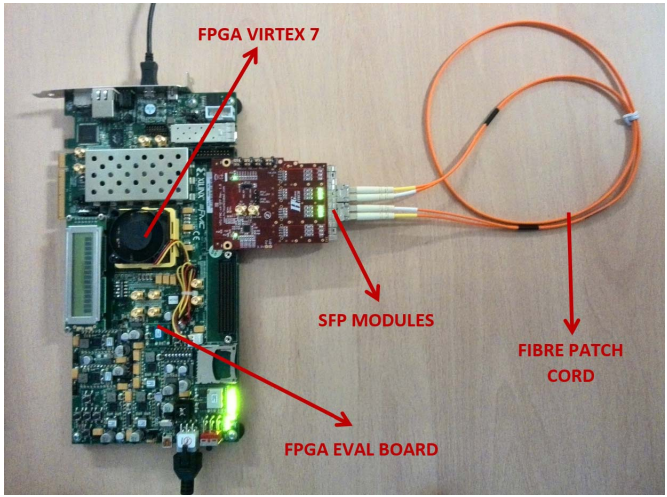

Fig. 15. Test setup photograph.

The FPGA design consists of two Ethernet interfaces with PHYsec function and two Ethernet frame generator modules connected to them. Each Ethernet interface contains the MAC module and the PHY (PCS and PMA blocks) including the encryption system explained along this paper. The scheme of the overall hardware system and the structure of the Ethernet interfaces are shown in Figs. 13 and 14, respectively. In Fig. 15, a photograph of the test setup is also shown.

The PHY is connected directly to the SFP modules, thanks to the FPGA SERDES circuit. In the MAC side, Ethernet interface is connected to the Ethernet frame generator to test the encrypted link with real traffic and verify that no frames are lost and no cyclic redundancy check (CRC) errors are produced during encryption.

It is important to remark that the initial PCS structure of this paper (without the encryption mechanism) parts from an implementation compatible with the standard. The PHYsec functionality has been developed and incorporated to this initial PCS sublayer.

Moreover, the final PCS structure, including PHYsec functionality, introduces an extra latency in the 8b10b TX datapath of 192 ns, in respect the baseline implementation, and approximately the same in the RX datapath. This extra latency is due to the new hardware modules added to PCS sublayer and shown in Fig. 3. For example, in the TX direction, 8b10b symbols have to traverse INSERT and CIPHER_OP_TX modules before being encoded. In the INSERT module, detailed in Fig. 6, the pipeline introduces a latency of 18 clock cycles, while the CIPHER_OP_TX operations (mapping, modulo-267 addition, and reverse-mapping) delays only take six clock cycles. The total latency introduced in TX datapath is 24 cycles at a clock rate of 125 MHz (8 ns of period), which is the operation frequency at which the system works.

### B. Encryption Results

The conclusions drawn from simulation and hardware debugging can be summarized in the following points.

1) Encryption/decryption works correctly and synchronously without harming data traffic or link establishment between Ethernet interfaces. Maximum data rate is achieved without frame losses or CRC errors. It has been checked, thanks to frame and CRC counters inside Ethernet frame generators. Traffic bursts were tested with a duration of $10^6$ frames with length of 1500 bytes. The throughput of the traffic was configured between 10% and 98% of the maximum line rate.

2) Encryption allows making the transmitted frames indecipherable. When encryption keys are different between transmitter and receiver, no valid frames are received in RX interface as it is impossible to detect the right 8b10b symbol flow.

3) Encryption makes indistinguishable a data traffic pattern from a continuous idle transmission, and then it is able to hide the pattern of Ethernet traffic from passive eavesdroppers.

For this last capability, it is interesting to monitor the signal waveforms at the input of the encoder and output of the decoder. Particularly, K control flag can give information about the transmission state. This flag is generated next to each 8b10b symbol by the PCS_TX_CTRL controller in the PCS sublayer. Both K control flag and 8b10b symbols are the input to the encoder and the output of the decoder. Each 8b10b symbol is a control or data one depending whether its K flag is "1" or "0," respectively.

If the encryption is not enabled, when transmitting no frames, K control flag pattern is a signal that switches continuously between "0" and "1." It is because when no frame is transmitted, idle ordered sets are continuously sent, and they are composed by two consecutive symbols, the control symbol /K28.5/ plus a data symbol (/D5.6/ or /D16.2/).

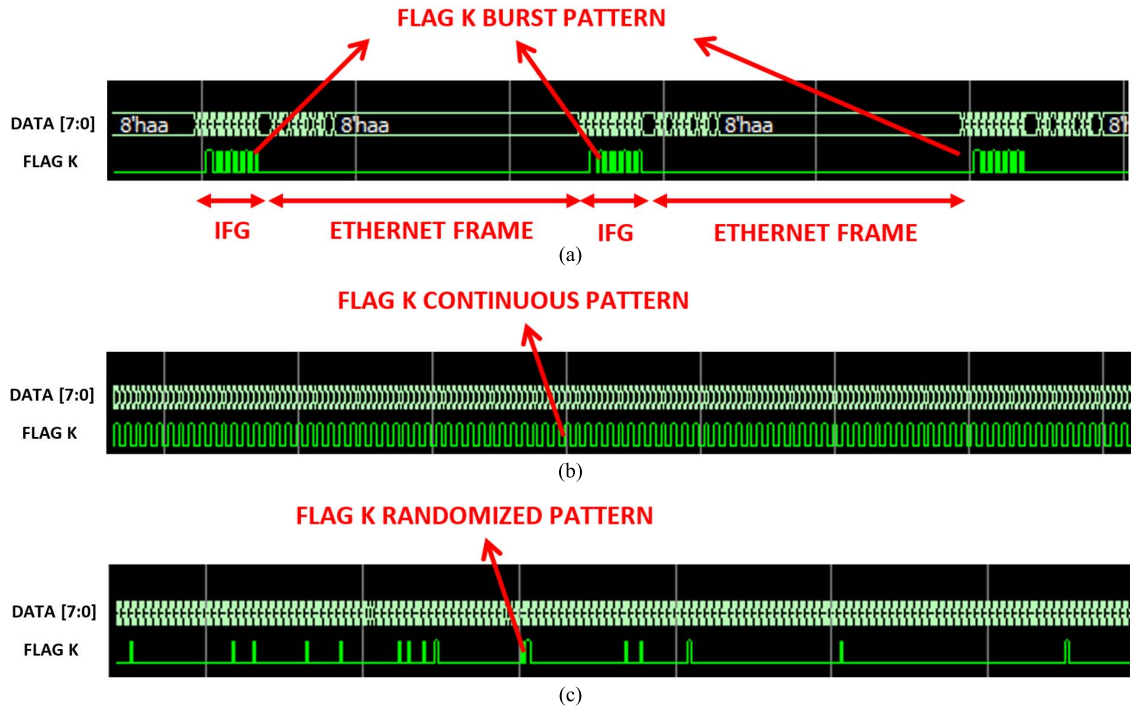

Fig. 16. (a) K flag pattern before encryption when transmitting an Ethernet frame burst. (b) K flag pattern before encryption when no Ethernet frame is transmitted. (c) K flag pattern after encyption regardless of the transmission or nontransmission of Ethernet frames.

TABLE II
FPGA Resources Used in Keystream Generator Submodules

| MODULE | Slice LUTs | Slice Registers | MULT cells |
|---|---|---|---|
| KEYSTREAM_GEN[1] | 10943 | 3629 | 144 |
| LFSR | 210 | 195 | 0 |
| STM_BANK | 6606 | 589 | 144 |
| MOD-267 | 4127 | 2845 | 0 |

[1]KEYSTREAM_GEN refers to the hardware resources of the Keystream Generator, including LFSR, STM_BANK and MOD-267 operation.

TABLE III
Comparison With Other Solutions

| | FPE [23] | This Work |
|---|---|---|
| Slice Registers | 11127 | 3629 |
| Slice LUTs | 16978 | 10943 |
| 18K Block RAMs | 77 | 0 |
| MULT cells | 0 | 144 |
| Slices[1] | 5636 | 3190 |
| Operation Freq. (MHz) | 125 | 125 |
| Encryption Rate (Mbps) | 1000 | 1000 |
| Encryption Rate/Slice (Kbps/Slice) | 177.4 | 313.4 |
| Power consumption (W) | 0.775 | 0.332 |

[1]Slices are estimated from the number of register and LUTs, assuming they are not packed together.

However, when transmitting Ethernet frames, the K flag pattern seems a burst pattern, as idle transmission only occurs in the IFG periods and only data symbols are transmitted between frame boundaries (setting K flag to zero).

When encryption is enabled, K flag pattern seems completely random in both situations, with or without Ethernet traffic being transmitted. This effect makes indistinguishable the data traffic pattern. In Fig. 16, simulation screenshot shows how the K flag pattern in the ciphertext [Fig. 16(c)] is randomized regardless of the traffic pattern of the plaintext [Fig. 16(a) and (b)].

### C. Implementation Results

Regarding the resources used by the proposed solution, in Table II, the main contribution for each block inside each keystream generator (KEYSTREAM GEN) is shown. Modulo 267 (MOD-267) operation and keystream generator bank (STM_BANK) take up the largest amount of FPGA resources.

In the case of multiplication (MULT) cells, 16 of them are necessary to implement the multiplications inside each STM_CELL, as the chaotic map implementation requires it. As there are nine STM cells per STM_BANK, then a total of 144 MULT cells are necessary for each keystream generator.

Moreover, a comparison in terms of hardware resources with the solution proposed in [23] is shown in Table III. The main difference between them is the number of block RAM (BRAM) and MULT cells. While this solution uses MULT cells, in [23], no cell of this type is used. However, the opposite happens with the number of BRAMs.

In addition, power consumption has been estimated in both solutions. First, postsynthesis simulation of both mechanisms has been performed to obtain the switching activity inter-change format files, where port and signal switching rates are recorded. With this information and thanks to the power estimation tools, provided by the FPGA manufacturer, the dynamic power figures have been obtained. The implementation of this paper achieves clearly better figures in *Encryption_Rate/Slice* and power consumption as shown in Table III.

TABLE IV
Chaotic Cell Comparison With Other Solutions

| | LM [39] | MLM [28] | Bernoulli [30] | Hennon [29] | This Work |
|---|---|---|---|---|---|
| Platform | Virtex 5 | Virtex 6 | Spartan 3E | Virtex 6 | Virtex 7 |
| Slice Registers | 64 | 160 | 108 | 64 | 65 |
| Slice LUTs | 129 | 643 | 575 | 1600 | 734 |
| MULT cells | 16 | 16 | 9 | 16 | 16 |
| Slices[1] | 49 | 128 | 342 | 275 | 131 |
| Max. Freq. (MHz) | 26.9 | 93 | 36.9 | 25.7 | 174 |
| Output width | 1 | 16 | 1 | 1 | 8 |
| Encryption Rate (Mbps) | 26.9 | 1488 | 7.38 | 25.7 | 1392 |
| Encryption Rate/ (Slice×Output width) (Mbps/(slice×bit)) | 0.21 | 0.73 | 0.02 | 0.1 | 1.32 |

[1]Slices are estimated from the number of register and LUTs, assuming they are not packed together.

Thanks to this comparison, it is possible to conclude that the structure proposed in this paper, based on a stream cipher next to a modulo operation, is more efficient than an FPE blockcipher working in CTR mode [23].

However, as the stream cipher structure is based on a chaotic map, it is also interesting to make a comparison of the presented chaotic cell with other chaotic solutions. In Table IV, a comparison in terms of hardware resources is made with other chaotic implementations. Particularly, solutions implementing the logistic map [39], modified logistic map [28], Bernoulli [30], and Hennon [29] maps are shown. While in Tables II and III, the overall hardware resources of this solution are shown, and in Table IV, only the hardware relative to the STM cell is taken into account.

As each implementation takes from its internal state different number of bits as output, the comparison has been made in terms of the *Encryption_rate* per slice and output bit. According to this, the proposed solution clearly achieves a better result.

## VII. Conclusion

As far as the authors are aware, this is the first time that a chaotic solution for encrypting 1000Base-X Ethernet physical layer has been proposed and developed. The new encryption function PHYsec consists of symmetric ciphering at PCS sublayer of the 8b10b symbol stream transmitted over an optical link. Encryption based on an original chaotic cipher has been tested with real Ethernet traffic and it has been concluded that the proposed system works correctly without harming data traffic or link establishment. In this paper, not only Ethernet frames are ciphered but also the data traffic patterns are masked. These features could improve the security at physical level with no throughput losses, zero space overhead, and low latency.

On the other hand, the proposed keystream generator module entails ratio throughput/resources better than existing FPE implementations. Moreover, other chaotic or nonchaotic secure stream ciphers could be compatible in the proposed scheme.

Finally, as the proposed PHYsec method is suitable for PCS sublayers using 8b10b encoding, the same idea could be extended to other high-speed standards based on this codification, such as Fiber Channel or peripheral component interconnect-express.

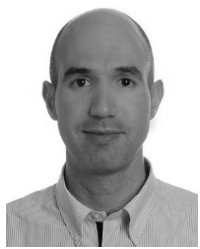

**Adrián Pérez-Resa** was born in San Sebastián, Spain. He received the M.Sc. degree in telecommunications engineering from the University of Zaragoza, Zaragoza, Spain, in 2005, where he is currently pursuing the Ph.D. degree with the Group of Electronic Design, Aragón Institute of Engineering Research.

He was a Research and Development Engineer with Telnet-RI, Zaragoza. He is currently a member of the Group of Electronic Design, Aragón Institute of Engineering Research, University of Zaragoza. His current research interests include high-speed communications and cryptography applications.

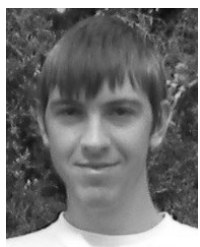

**Miguel Garcia-Bosque** was born in Zaragoza, Spain. He received the B.Sc. and M.Sc. degrees in physics from the University of Zaragoza, Zaragoza, in 2014 and 2015, respectively.

He is currently a member of the Group of Electronic Design, Aragón Institute of Engineering Research, University of Zaragoza. His current research interests include chaos theory and cryptography algorithms.

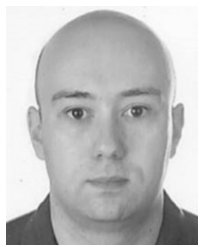

**Carlos Sánchez-Azqueta** was born in Zaragoza, Spain. He received the B.Sc., M.Sc., and Ph.D. degrees in physics from the University of Zaragoza, Zaragoza, in 2006, 2010, and 2012, respectively, and the Dipl.-Ing. degree in electronic engineering from the Complutense University of Madrid, Madrid, Spain, in 2009.

He is currently a member of the Group of Electronic Design, Aragón Institute of Engineering Research, University of Zaragoza. His current research interests include mixed-signal integrated circuits, high-frequency analog communications, and cryptography applications.

**Santiago Celma** was born in Zaragoza, Spain. He received the B.Sc., M.Sc., and Ph.D. degrees in physics from the University of Zaragoza, Zaragoza, in 1987, 1989, and 1993, respectively.

He is currently a Full Professor with the Group of Electronic Design, Aragon Institute of Engineering Research, University of Zaragoza. He is a Principal Investigator for more than 30 national and international research projects. He has co-authored more than 100 technical papers, 300 international conference contributions, and four technical books. He holds four patents. His current research interests include circuit theory, mixed-signal integrated circuits, high-frequency communication circuits, wireless sensor networks, and cryptography for secure communications.